\documentclass[11pt]{article}
\usepackage[utf8]{inputenc}  
\usepackage[T1]{fontenc}     
\usepackage[british]{babel}  
\usepackage[colorlinks=true]{hyperref}
\hypersetup{colorlinks=true,linkcolor=teal,citecolor=blue,urlcolor=blue}
\usepackage[margin=10pt,font=small,format=hang,figurename=Fig. ,indention=-.55cm,labelfont={bf,sf,sc}]{caption}
\usepackage{graphicx,epstopdf}
\usepackage{amsmath}
\usepackage{amssymb}
\usepackage{hyperref}
\usepackage{latexsym}
\usepackage{color}
\usepackage{subfigure}

\usepackage{pst-all}
\usepackage{ifpdf}
\usepackage{multimedia}
\usepackage{lmodern}
\usepackage{cite}
\usepackage{dsfont}
\usepackage{bbm}
\usepackage{lipsum}
\usepackage{url}
\usepackage{textcomp}
\usepackage{bbold}
\usepackage{mathtools}
\usepackage{nccmath}
 \usepackage[normalem]{ulem}
\usepackage{ifpdf}
\ifpdf
\else
\fi
  \hypersetup{pdfstartview=XYZ}
\newcommand{\bra}{\begin{array}}
\newcommand{\era}{\end{array}}
\newcommand{\beq}{\begin{equation}}
\newcommand{\eeq}{\end{equation}}
\newcommand{\beqar}{\begin{eqnarray}}
\newcommand{\eeqar}{\end{eqnarray}}

\def\BC{\bb C}
\def\_\BC{\bbi C}

\def\( {\left(}
\def\) {\right)}
\def\[ {\left[}
\def\] {\right]}
\def\no2 {{\textstyle{n\over 2}}}

\newcommand{\lb}{\label}
\begin{document}
\thispagestyle{empty}
\begin{center}

\vspace{1.8cm}

 {\Large {\bf Tunable Dynamics of a Dipolar Quantum Battery: Role of Spin-Spin Interactions and Coherence}}\\


\vspace{1.5cm} {\bf J. Ramya Parkavi}{\footnote {email: {\sf ramyaparkji@gmail.com}}}, {\bf R. Muthuganesan}{\footnote{email: {\sf
rajendramuthu@gmail.com}}} and {\bf V.K. Chandrasekar}{\footnote {email: {\sf chandru25nld@gmail.com}}}  \\
\vspace{0.5cm}
{\it $^1$Department of Physics, School of Sciences, SRM Institute of Science and Technology, Tiruchirappalli 621 105, Tamil Nadu, India}\\ [0.5em]
{\it $^2$Department of Physics and Nanotechnology, College of Engineering and Technology, SRM
Institute of Science and Technology, Kattankulathur 603 202, Tamil Nadu, India}\\ [0.5em]
{\it $^3$Department of Physics, Centre for Nonlinear Science and Engineering, School of Electrical and Electronics Engineering, SASTRA Deemed University, Thanjavur, 613 402, Tamil Nadu, India}\\ 

\end{center}
\baselineskip=18pt
\medskip
\vspace{3cm}
\begin{abstract}
This study explores the energy storage dynamics of a quantum battery (QB) modeled using a dipolar spin system with Dzyaloshinskii–Moriya (DM) interaction. We examine the performance of this system in terms of ergotropy, instantaneous power, capacity, and quantum coherence using a two-qubit model. By solving the system's time evolution under cyclic unitary processes, we analyze how external parameters such as temperature, magnetic field, and DM interaction influence the charging behavior and quantum resources of the battery. The findings demonstrate that quantum coherence and DM interaction significantly enhance the energy storage efficiency and power output of the quantum battery, offering promising strategies for designing high-performance quantum energy storage devices. Furthermore, we investigate the performance of quantum battery under the influence of a common dephasing environment, which  limits the long-term work-extraction capability of dipolar quantum batteries.

\end{abstract}
\vspace{1cm}
~~~~~~~~~\noindent{\it Keywords}: Quantum Battery; Dzyaloshinsky-Moriya (DM)  interaction; Dipolar system; Quantum coherence;
\newpage
 \renewcommand{\thefootnote}{*}
\section{Introduction}
With the ever-shrinking scale of electronic devices, quantum phenomena are becoming increasingly significant, influencing areas from fundamental physics to cutting-edge technological applications. These quantum effects can offer distinct advantages, enabling the development of devices that outperform their classical counterparts in speed, efficiency, and functionality \cite{quantumdevices1, quantumdevices2, quantumdevices3, quantumdevices4}. One such promising technology is the quantum battery (QB), a quantum storage system designed to store and deliver energy using the principles of quantum mechanics. The concept of a quantum battery was first formalized by Alicki and Fannes in 2013, who introduced it within an information-theoretic framework by quantifying the maximum amount of energy that can be extracted from a quantum system through unitary evolution \cite{Alicki}. Since then, quantum batteries have garnered growing interest due to their potential to enable ultra-fast charging, high energy density, miniaturization, and enhanced energy extraction efficiency, which are critical features for future nanoscale and quantum technologies \cite{reviewarticle}. These characteristics make QBs highly suitable for applications in quantum computing, quantum sensing, and next-generation portable electronics, where conventional energy systems fall short. It is important to note the distinction between a quantum battery and other quantum-inspired storage devices, such as quantum capacitors. A quantum capacitor stores energy through electrostatic charge accumulation in close analogy with its classical counterpart, and its operation does not inherently require quantum coherence. By contrast, quantum batteries exploit quantum coherence and correlations to store and release energy, offering enhanced charging power and extractable work beyond what is achievable in capacitor-based models \cite{reviewarticle, QBelectroarticle, QCbook}. However, the classical version of the battery also includes devices that use different materials \cite{chembattery1}, \cite{chembattery2}.

Ferraro \textit{et al.} \cite{firstQBsolidstate} proposed the first solid-state model of a quantum battery, sparking a surge of interest in diverse physical implementations. Quantum-battery prototypes have since been realised or theoretically designed in cavity-optomechanical platforms \cite{cavity1,cavity2, cavity3}, spin ensembles \cite{NMR,spininteraction}, superconducting circuits \cite{supercond1,supercond2}, Dicke-model systems \cite{Dickemodel1,Dickemodel2}, and photonic architectures \cite{photonicQB1,photonicQB2}. Within spin systems, an XYZ Heisenberg battery under an external magnetic field was studied in Ref.~\cite{indrajith}, demonstrating that field tuning critically controls the stored energy and charging power. A three-spin XXZ battery revealed that ferromagnetic versus antiferromagnetic initial states lead to distinct optimal charging strategies \cite{threespinbattery}. More recently, precise parameter engineering in superconducting quantum batteries has been shown to boost the extractable work dramatically \cite{superconducting}. Yet, designing a model that delivers a clear quantum advantage while remaining experimentally accessible is still challenging. In this context, we investigate a dipolar quantum battery whose long-range dipole–dipole and Dzyaloshinskii–Moriya interactions give rise to rich charging and energy-storage dynamics. Dipolar systems are experimentally available in settings ranging from molecular magnets to ultracold Rydberg arrays and continue to attract attention for both their technological promise and their intriguing quantum many-body behaviour \cite{dipolarsystem1,dipolarsystem2,Reis,dipolarspinsystem}. 


In recent developments, quantum measurements have emerged as fundamental tools in governing the behavior of quantum batteries, enabling the design of more controllable and efficient energy storage systems \cite{Nonequilibrium, Localprojective, correlatedcharges}. As a result, considerable effort has been directed toward developing protocols that maximize work extraction from quantum batteries, particularly by harnessing quantum entanglement and quantum correlations \cite{firstQBsolidstate, entangleQB2, entangleQB3, entangleQB4}. However, as demonstrated in Ref. \cite{entanglementcharging}, entanglement is not always the most effective resource for charging quantum batteries. The study examined two-cell and three-cell quantum batteries and revealed that coherence, rather than entanglement, can play a more critical role depending on the system. Coherence generation was identified as a key resource for achieving optimal charging performance. Likewise, other works \cite{superconducting, CoherenceQB1, CoherenceQB2} have also emphasized the advantages of quantum coherence in enhancing battery performance. In this context, our work explores a coherence-driven quantum battery model, highlighting the pivotal role of quantum coherence in improving energy storage efficiency and charging dynamics.

This paper explores dipolar spin systems with Dzyaloshinskii–Moriya (DM) interaction, exploites their quantum coherence to extract ergotropy and enhance energy capacity. We further investigate the role of external magnetic fields and other system parameters in shaping the performance of a dipolar quantum battery (QB). For this model, we introduce an effective parameter that characterizes the maximum energy storage capacity of the battery. Our findings underscore the significance of tailoring the Hamiltonian parameters and highlight the critical role of spin–spin interactions in optimizing the energy storage mechanism of quantum batteries. It is worth noting that while recent works have investigated the role of DM interaction in Heisenberg spin-chain quantum batteries \cite{Refereearticle2, Refereearticle1}, our study differs in several key aspects. In particular, we focus on a dipolar spin-system model, where long-range dipole–dipole couplings coexist with DM interaction, offering a richer dynamical landscape than short-range Heisenberg models. Moreover, our analysis explicitly employs ergotropy and coherence as performance quantifiers, thereby establishing a direct link between quantum resources and extractable work—an aspect that is not fully addressed in earlier studies. Finally, we extend the discussion to open-system dynamics by including the effect of a common dephasing environment, providing practical insights into the robustness and limitations of dipolar quantum batteries under realistic conditions.

We organize the article as follows: In Section ~\ref{sec2}, we review the mathematical formulations of ergotropy, instantaneous power, capacity, and quantum coherence. Section~\ref{Sec3} presents the model of the dipolar spin system, including the Dzyaloshinskii–Moriya (DM) interaction and other relevant interactions considered in this study. In Section~\ref{Sec4}, we analyze the effects of key system parameters-temperature, magnetic field, and DM interaction—on the performance indicators of the quantum battery, supported by numerical simulations and plots. This section also includes a qualitative discussion of the observed trends and insights into the battery's behavior. We examine the impact of a dephasing environment on the battery-charger system and provide a brief overview of the experimental realization in Section~\ref{dephasesec}. Finally, Section~\ref{concl} provides a summary of our main findings and concluding remarks.

\section{Quantum Battery}\label{sec2}
To design an efficient quantum battery, we model an interacting dipolar spin system with Dzyaloshinskii-Moriya (DM) interaction as a quantum battery. In general, the storable energy Sland ergotropy serve as key figures of merit for evaluating the performance and effectiveness of the quantum battery \cite{ergotropybasepaper}. The spectral decomposition of the quantum battery, given by $$\mathcal{H}_{\mathcal{B}}=\sum_{i=1}^4\lambda_i|\varphi_i\rangle \langle \varphi_i|$$ 
is instrumental in understanding energy variations and work extraction during cyclic unitary processes. Here, $\lambda_i$ and $|\varphi_i\rangle$  are the eigenvalues and eigenvectors of $\mathcal{H}_{\mathcal{B}}$ respectively. This decomposition suggests that the Hamiltonian of the quantum battery must return to its initial condition at the end of cyclic unitary processes.  At time $t=0$, the initial state of the battery is assumed to be either the ground state of $\mathcal{H}_{\mathcal{B}}$  at absolute zero or the Gibbs state at a finite temperature, as calculated in Equation (\ref{thermal}). The time evolution of the $\varrho$ is governed by the Liouville–von Neumann equation:
\begin{align}
\frac{d}{dt}\varrho(t)=-i \left[\mathcal{H}_{\mathcal{B}}+\mathcal{H}_c(t), \varrho(t) \right]
\end{align}
where $\mathcal{H}_c(t)$ is the time-dependent charging during the time $[0, t]$ defined via Pauli X-gate. The charging Hamiltonian is given as 
\begin{align}
\mathcal{H}_c(t)=\Omega(\sigma_x\otimes \mathds{1}+\mathds{1}\otimes \sigma_x),
\end{align}
where $\Omega$ represents the strength (frequency) of the external field that injects energy into the dipolar quantum battery, $\sigma_x$ is the Pauli matrices and $\mathds{1}$ is the identity matrix.
In a closed system, the charging process can be described by the unitary time evolution operator
\begin{align}
\mathcal{U}_c(t)=\text{exp}[-i\mathcal{H}_c t].
\end{align}
 The explicit form of the unitary operator corresponding to the X-gate charging process is
 \begin{align}
\mathcal{U}_c(t) = \begin{pmatrix}
 r & s & s & q \\
 s & r & q & s \\
 s & q  & r & s \\
q & s & s &  r
\end{pmatrix},
\label{unitray}
\end{align}
where $r=\text{cos}^2(\Omega t)$, $q=-\text{sin}^2(\Omega t)$ and $s=-\text{sin}(2\Omega t)/2$. After extracting the maximum amount of work in the unitary process, the system reaches a passive state $\sigma$ which is diagonal in the Hamiltonian’s eigenbasis. 

The work extracted during this process, known as ergotropy, is given by
\begin{align}
W(t)=\text{Tr}[\varrho(t) \mathcal{H}_{\mathcal{B}}]-\text{Tr}[\varrho(T) \mathcal{H}_{\mathcal{B}}]
\end{align}
where $\varrho(T)$ is the thermal density matrix at temperature $T$ and $\varrho(t)$ is the density matrix of the system at time $t$, calculated from the unitary transformation as
\begin{align}
\varrho(t)=\mathcal{U}_c(t) \varrho(T) \mathcal{U}_c(t)^{\dagger}.
\end{align}


The instantaneous power of a quantum battery is defined as
\begin{align}
\mathcal{P}(t) = \frac{dW(t)}{dt},
\end{align}
where $W(t)$ is the work done at time $t$. 
Another key performance metric of a quantum battery is its capacity, denoted by $\mathit{K}$ \cite{capacity}.  Without solving complex dynamical equations, this measure provides valuable insight into the energy storage potential of the system:
\begin{align}
\mathit{K} = \text{Tr}[H_{\mathcal{B}} \hat{\varrho}_\uparrow] - \text{Tr}[H_{\mathcal{B}} \hat{\varrho}_\downarrow],
\end{align}
where  $\hat{\varrho}_\downarrow=|0^{\otimes N} \rangle \langle 0^{\otimes N} |$ and  $\hat{\varrho}_\uparrow=|1^{\otimes N} \rangle \langle 1^{\otimes N} |$ represent the ground and excited states of the $N$-partite quantum battery, respectively. For our case, we consider a two-qubit battery, i.e., $N=2$. This metric directly quantifies the energy gap between the fully charged and completely discharged states. Importantly, since it does not require time-dependent optimization, $\mathit{K}$ is applicable to both closed and open quantum systems, making it a practical and efficient tool for evaluating quantum battery performance.

\par Recent investigations have explored the role of quantum coherence in energy storage systems \cite{CoherenceQB1, CoherenceQB2}. Accordingly, we quantify the coherence of the quantum battery using the $l_1$-norm of coherence, defined as:
\begin{align}
C_{l_1}(t) = \frac{1}{C_{\text{max}}} \sum_{i \neq j} |\varrho_{ij}(t)|,
\label{l1norm}
\end{align}
where the quantity $C_{max}$ represents the maximum possible coherence for the system. For a two-qubit state, $C_{max}=
3$, which corresponds to the maximally coherent state  $|\psi\rangle=(1/2)(|\uparrow \rangle + |\downarrow \rangle)(|\uparrow \rangle + |\downarrow \rangle )$. By normalizing the definition of the $l_{1}$ norm of coherence \cite{l1normalization1, l1normalization2}, we define the aforementioned quantity such that $0\leq C_{l_1}(t) \leq 1$. Then, one can examine the QB's ``quantum'' nature from Equation (\ref{l1norm}). In our formalism, the ergotropy corresponds to the maximum usable energy that can be extracted from the spin system, much like the dischargeable energy in a conventional battery. However, unlike classical storage, this extractable work arises from quantum coherence and spin–spin correlations, which act as resources enabling faster charging and enhanced power output. Thus, the theoretical measures introduced here have a direct physical interpretation in terms of the effective energy storage and release capabilities of the dipolar quantum battery.


\section{The model and thermalization}\label{Sec3}
\begin{figure}[!htbp]
	\centering
	\includegraphics[width=0.75\textwidth]{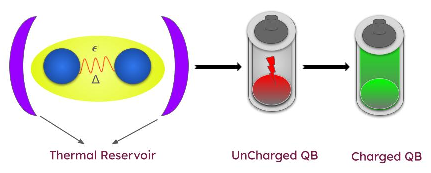}
	\caption{Schematic representation of dipolar quantum battery (QB)}
	\label{Fig0}
\end{figure}

To construct the quantum battery as shown in Figure (\ref{Fig0}), we consider a two-spin system coupled via dipolar \cite{Reis, dipolartele} and Dzyaloshinskii–Moriya (DM) interactions \cite{Dzyaloshinskii, Moriya}. Let $\vec{S}_{1} = (S_{1}^{x}, S_{1}^{y}, S_{1}^{z})$ and $\vec{S}_{2} = (S_{2}^{x}, S_{2}^{y}, S_{2}^{z})$ denote the spin-$\tfrac{1}{2}$ operators of the first and second particles, respectively, where $S_{i}^{\alpha} = \tfrac{1}{2}\sigma_{i}^{\alpha}$ ($\alpha = x, y, z$) and $\sigma_{i}^{\alpha}$ are the Pauli matrices acting on spin $i$. With this convention, the dipolar interaction Hamiltonian can be written as \cite{spinmodeleqn2, spinmodeleqn5, spinmodeleqn1, spinmodeleqn4, spinmodeleqn3}
\begin{equation}
H_{B} = -\tfrac{1}{3}\,\vec{S}_{1}\cdot\overleftrightarrow{\mathcal{D}}\cdot\vec{S}_{2} 
+ \vec{D}\cdot(\vec{S}_{1}\times\vec{S}_{2}) 
+ \vec{B}\cdot(\vec{S}_{1} + \vec{S}_{2}),
\label{ham}
\end{equation}
where $\overleftrightarrow{\mathcal{D}}$ is the dipolar interaction tensor, $\vec{D}$ represents the DM interaction vector, and $\vec{B}$ is the homogeneous external magnetic field. The tensor
\begin{align}
\overleftrightarrow{\mathcal{D}}=\text{diag}(\Delta-3\epsilon, \Delta+3\epsilon,-2\Delta ) \nonumber 
\end{align}
is a traceless diagonal tensor, with $\Delta$ and $\epsilon$ are the dipolar coupling constants between the spins. The vector  $\vec{D}$ denotes the DM coupling vector and $\vec{B}$ is homogeneous magnetic field. In our setup, both the DM coupling and the magnetic field are aligned along the $z-$ axis. The sign of $\Delta$ indicates the orientation of the spin: if $\Delta< 0$, the spin is in the $x-y$-plane, whereas if $\Delta> 0$, the spin is directed along the z-axis. The dipolar interaction arises due to the magnetic moment of a spin on another spin located at the nearest site and the DM interaction due to spin-orbit coupling. The dipolar interaction arises from the magnetic moment of one spin influencing a neighboring spin, while the DM interaction originates from spin–orbit coupling. Here, the dipolar coupling constants arise microscopically from the direct magnetic dipole-dipole interaction between localized spin moments, and are routinely characterized in systems such as NMR spin ensembles, molecular magnets, and ultracold Rydberg arrays \cite{Reis}.
 
In the standard two-qubit computational basis $\{ |00\rangle,|01\rangle,|10\rangle,|11\rangle \} $, the Hamiltonian $\mathcal{H}_{\mathcal{B}}$ takes the form
\begin{align}
\mathcal{H}_{\mathcal{B}} = \begin{pmatrix}
 \frac{1}{6}(\Delta+3B) & 0 & 0 & \frac{\epsilon}{2} \\
 0 & -\frac{\Delta}{6} & -\frac{\Delta}{6} + \mathrm{i} \frac{D}{2} & 0 \\
 0 & -\frac{\Delta}{6}-\mathrm{i} \frac{D}{2}  & -\frac{\Delta}{6} & 0 \\
 \epsilon/2 & 0 & 0 &  \frac{1}{6}(\Delta-3B)
\end{pmatrix}.
\label{ham2}
\end{align}
The eigenvalues and corresponding eigenvectors of the Hamiltonian (\ref{ham2}) are computed as  
\begin{subequations}\lb{eigen}
	\begin{align}
\lb{eq1} \lambda_{1}=  \frac{1}{6}(\Delta + 3\eta), ~& ~~~~~~~~~~~~~~~~~~~~~~~~~~~~~~~~ \vert \varphi_{1}\rangle= \Gamma_{+}\left(\frac{B+ \eta}{\epsilon}\vert 00 \rangle + \vert 11 \rangle\right),~~~~~~~~~~~~\\
   \lb{eq2} \lambda_{2}= \frac{1}{6}(-\Delta + \chi), & ~~~~~~~~~~~~~~~~~~~~~~~~~~~~~~~~\vert \varphi_{2}\rangle= \Lambda_{+}\left(\frac{ \chi}{6i D-\Delta}\vert 01 \rangle + \vert 10 \rangle\right), ~~~~~~~~~~~~\\
\lb{eq3} \lambda_{3}= \frac{1}{6}(-\Delta -\chi), & ~~~~~~~~~~~~~~~~~~~~~~~~~~~~~~~~\vert \varphi_{3}\rangle= \Lambda_{-}\left(\frac{-\chi}{6i D-\Delta}\vert 01 \rangle + \vert 10 \rangle\right),~~~~~~~~~~~~\\
   \lb{eq4} \lambda_{4}=  \frac{1}{6}(\Delta - 3\eta), ~& ~~~~~~~~~~~~~~~~~~~~~~~~~~~~~~~~ \vert \varphi_{1,4}\rangle= \Gamma_{-}\left(\frac{B - \eta}{\epsilon}\vert 00 \rangle + \vert 11 \rangle\right),~~~~~~~~
\end{align}
\end{subequations}
where $\eta=\sqrt{B^2+\epsilon^2}$ and $\chi=\sqrt{\Delta^2+9D^2}$. The normalization constants are 
\begin{align}
 \Gamma_{\pm}=\left(1+\frac{(B\pm \eta)^2}{\epsilon^2} \right)^{-1/2} ~~~~~~~~~ \Lambda_{\pm}=\left(1+ \frac{\Delta^2 \pm 9D^2}{3\mathrm{i}D-\Delta}\right)^{-1/2}. \nonumber
\end{align}

At the thermal equilibrium, the thermal state of the battery is calculated  as
\beq\lb{partition1}
\varrho(T)=\frac{1}{\mathcal{Z}}\exp{\left(-\beta \mathcal{H}_{\mathcal{B}}\right)}=\frac{1}{\mathcal{Z}}\sum_{i=1}^4 p_i \vert \varphi_{i}\rangle \langle \varphi_{i}\vert,
\eeq
where ${\mathcal{Z}} = \text{Tr}\exp\left(-\beta \mathcal{H}_{\mathcal{B}}\right)$ is the canonical partition function of the system and $p_i$ are the eigenvalues of $\varrho(T)$. The parameter $\beta = 1/(k_BT)$ represents the inverse temperature, with $k_B$ being the Boltzmann constant, which is set to unity for simplicity. The thermal density matrix $\varrho(T)$ can be written as
\begin{align}
\varrho(T) = \frac{1}{\mathcal{Z}}\begin{pmatrix}
 \varrho_{11} & 0 & 0 & \varrho_{14} \\
 0 & \varrho_{22} & \varrho_{23} & 0 \\
 0 & \varrho_{23}^*  & \varrho_{22} & 0 \\
\varrho_{14} & 0 & 0 &  \varrho_{44}
\end{pmatrix},
\label{thermal}
\end{align}
where are the matrix elements 
\begin{align}
\varrho_{11}=\mathrm{e}^{\frac{-\beta \Delta}{6}}\left[ \text{cosh}\left( \frac{\eta \beta}{2}\right)-\frac{B}{\eta}\text{sinh}\left( \frac{\eta \beta}{2}\right)\right],~~~~~~~~~~~\varrho_{22}=\varrho_{33}=\mathrm{e}^{\frac{\beta \Delta}{6}}   \text{cosh}\left( \frac{\chi \beta}{6}\right), \nonumber \\
\varrho_{44}=\mathrm{e}^{\frac{-\beta \Delta}{6}}\left[ \text{cosh}\left( \frac{\eta \beta}{2}\right)+\frac{B}{\eta}\text{sinh}\left( \frac{\eta \beta}{2}\right)\right], ~~~~~~~~~~\varrho_{14}=-\frac{\epsilon}{\eta}\mathrm{e}^{\frac{-\beta \Delta}{6}} \text{sinh}\left(\frac{\eta \beta}{2} \right), \nonumber \\
\varrho_{23}=\frac{\Delta-3\mathrm{i}D}{\chi} \mathrm{e}^{\frac{\beta \Delta}{6}} \text{sinh}\left(\frac{\chi \beta}{6} \right). ~~~~~~~~~~~~~~~~~~~~~~~ \nonumber
\end{align}
and the partition function is $\mathcal{Z}=2\mathrm{e}^{-\Delta \beta/6} \text{cosh}\left( \frac{\eta \beta}{2}\right)+2\mathrm{e}^{\Delta \beta/6} \text{cosh}\left( \frac{\chi \beta}{6}\right)$.

\begin{figure}[t]
	\centering
	\includegraphics[width=0.9\textwidth]{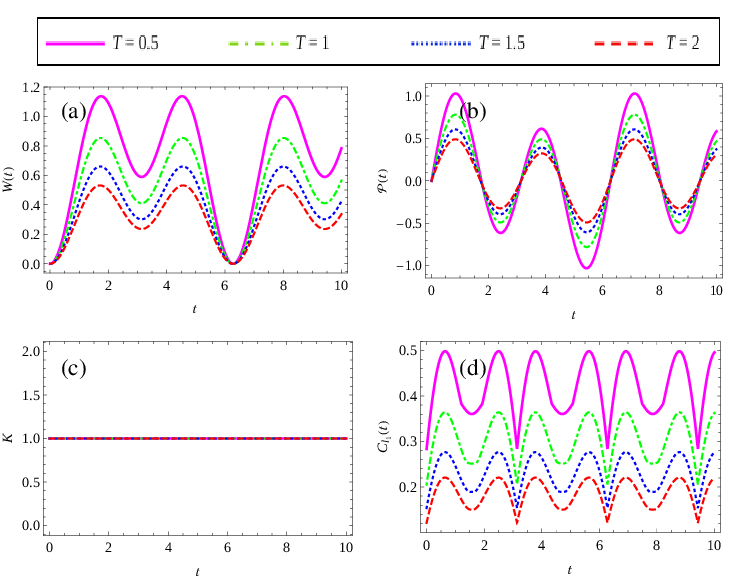}
	\caption{Time evolution of (a) ergotropy $W(t)$, (b) power $\mathcal{P}(t)$, (c) capacity $\mathit{K}$ and (d) $l_{1}$-norm of coherence $C_{l_1}(t)$ at different temperatures $T=0.5$, $1$, $1.5$, \text{and} $2$. Here, $\epsilon=\Delta=2$ are the dipolar coupling strengths, $D=1$ is the DM coupling, $B=1$ and $\Omega=1$ are the strengths of the external magnetic fields.}
	\label{Fig1}
\end{figure}
\section{Results and Discussions}\label{Sec4}
We present a detailed analysis of a quantum battery composed of a two-qubit system interacting via dipolar and Dzyaloshinskii–Moriya (DM) interactions under the influence of a homogeneous magnetic field. To evaluate the performance of this quantum battery, we employ key quantifiers such as ergotropy $W(t)$, instantaneous power $\mathcal{P}(t)$, energy capacity $\mathit{K}$, and the $l_1-$ norm of coherence $C_{l_1}(t)$.

In Figure (\ref{Fig1}a), we illustrate the time evolution of ergotropy for different temperatures. Ergotropy, which quantifies the maximum extractable work from a quantum system via a unitary transformation (relative to a passive state), displays an oscillatory behavior, reflecting periodic energy fluctuations within the system. As time progresses, we observe a sharp increase in ergotropy, indicating that the system is moving away from thermal equilibrium and accumulating usable energy. This is followed by a decay phase, forming a repeating cycle. Importantly, the magnitude of ergotropy, denoted as $W(t)$, diminishes with increasing temperature. This behavior suggests that lower temperatures significantly enhance the battery's performance to store and deliver usable energy, making the system more efficient for quantum energy storage and harvesting applications.

Figure (\ref{Fig1}b) shows the time evolution of power, $\mathcal{P}(t)$, which is defined as the time derivative of the stored energy and quantifies the rate at which energy is either delivered to or extracted from the battery. Similar to the behavior observed for ergotropy in Figure (\ref{Fig1}a), $\mathcal{P}(t)$ exhibits oscillatory and periodic variations over time. The peaks in the positive region of the power curve correspond to periods of maximum charging, indicating the most efficient energy input into the battery. Conversely, negative values of power signify discharging phases, where energy is being released from the battery. In essence, positive power denotes active charging of the quantum battery, while negative power reflects intervals of energy dissipation or extraction. In Figure (\ref{Fig1}c), we present the battery capacity as a function of time, which reveals that the capacity remains constant over time. Furthermore, we observe that the battery capacity remains constant regardless of temperature. Figure (\ref{Fig1}d) examines the $l_1 - $norm of coherence $C_{l_1}(t)$ as a function of time for various values of temperature. Similar to the behavior observed in Figure (\ref{Fig1}d), $C_{l_1}(t)$  shows oscillatory behavior as time evolves. In a similar manner, ergotropy and power, the amplitude of the oscillations in coherence slightly declines with increasing $T$, despite the frequency of oscillation increasing. In general, the thermal ﬂuctuation is always degrading the quantum resources and is clearly seen for higher values of temperature. We also observe that the increasing
temperature values tend to reduce the amount of quantum coherence in the system. However, at low temperature, coherence is maximum which reflects the possibility of achieving maximum extractable work from the considered system at low temperatures.
\begin{figure}[t]
	\centering
	\includegraphics[width=0.9\textwidth]{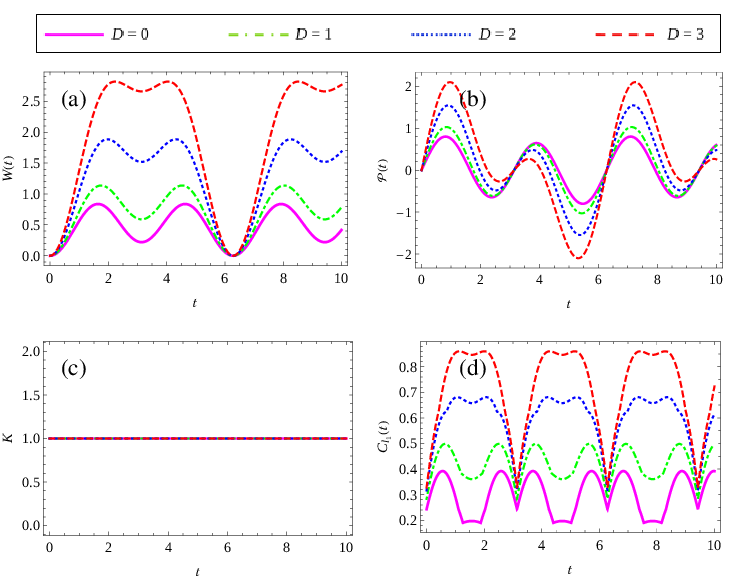}
	\caption{(a) Ergotropy $W(t)$, (b) power $\mathcal{P}(t)$, (c) capacity $\mathit{K}$ and (d) $l_{1}$-norm of coherence $C_{l_1}(t)$ evolve with time $t$ for varying DM interaction strengths, $D=0$, $1$, $2$, \text{and} $3$. Other parametric values are $\epsilon=\Delta=2$, $B=1$, $T=0.5$, and $\Omega=1$.}
	\label{Fig2}
\end{figure}

Figure (\ref{Fig2}) explores how the Dzyaloshinskii–Moriya (DM) interaction parameter $D$ influences the performance metrics of a dipolar quantum battery. Four key quantities are plotted as a function of time $t$ for different values of $D=0, 1, 2 ~\&~ 3$. In Figure (\ref{Fig2}a), we present the time evolution of ergotropy  $W(t)$  and observe that the amplitude of its oscillations increases with the strength of the DM interaction. This behavior suggests that DM interaction significantly enhances the energy extraction capability of the quantum battery. Notably, when  $D=0$ (pink solid curve), the ergotropy remains minimal, whereas higher values of $D$  lead to a substantial increase in its magnitude. Furthermore, previous studies on the thermal quantum correlations of this physical system \cite{dipolarspinsystem} revealed that DM interaction also enhances quantum correlations. This indicates a possible connection between increased quantum correlations and improved quantum battery performance, suggesting that stronger DM interaction may serve as a resource to boost the efficiency of energy storage and extraction in such quantum systems.

The instantaneous power is plotted in Figure (\ref{Fig2}b). The plot shows that as the DM interaction parameter $D$ increases, the oscillatory amplitude of the power  $\mathcal{P}(t)$ also increases in both the positive and negative directions. This behavior reflects more dynamic charging and discharging cycles, with higher peak power output. In essence, the quantum battery charges and discharges more rapidly when subjected to stronger DM interactions, indicating enhanced energy transfer rates. In contrast, the energy capacity, shown in Figure (\ref{Fig2}c), remains nearly constant over time across all values of  $D$. 
\begin{figure}[t]
	\centering
	\includegraphics[width=0.9\textwidth]{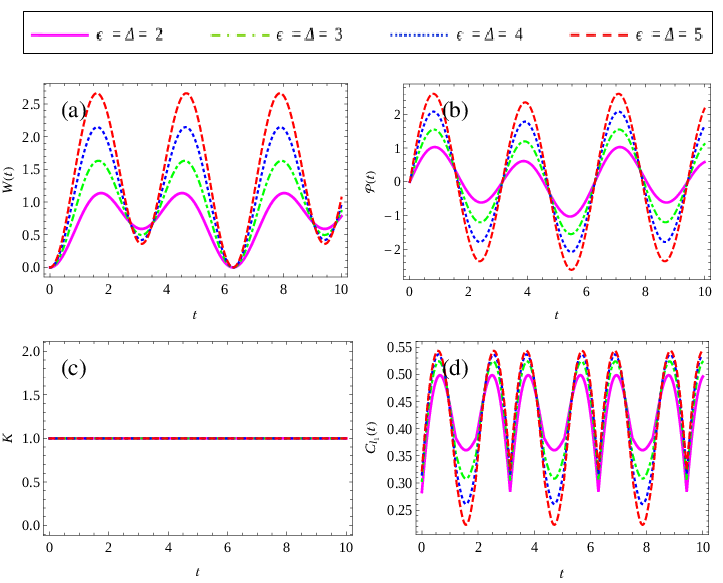}
	\caption{The temporal evolution of (a) ergotropy $W(t)$, (b) power $\mathcal{P}(t)$, (c) capacity $\mathit{K}$ and (d) $l_{1}$-norm of coherence $C_{l_1}(t)$ as a function of time for different dipolar coupling constants $\epsilon=\Delta=2$, $3$, $4$, and $5$. Herein, $T=0.5$ is the temperature, $D=1$ is the DM interaction,  $B=1$, and $\Omega=1$ are the applied fields.}
	\label{Energy factor}
\end{figure}
Figure (\ref{Fig2}d) displays the  $l_1-$norm of coherence $C_{l_1}(t)$ which quantifies the amount of quantum coherence in the system. As $D$ increases, a clear enhancement in coherence is observed. This is significant, as quantum coherence has been shown to be a key resource in boosting quantum battery performance. This behavior suggests that lower temperatures significantly enhance the battery’s performance in terms of extractable work (ergotropy) and coherence [Figures (2a), (2b), and (2d)], while the capacity $K$ remains independent of temperature as shown in Figure (2c). The observed correlation between coherence, ergotropy, and power implies that DM interaction not only induces stronger spin–spin correlations but also facilitates coherence generation, thereby improving the overall functionality of the battery. These results emphasize that engineering spin interactions through DM coupling provides a powerful approach to optimizing quantum energy storage systems. By tuning $D$, one can simultaneously enhance work extraction, increase charging/discharging efficiency, and maintain coherence-essential features for high-performance quantum batteries.

Figure \ref{Energy factor} illustrates the time evolution of four key physical quantities that characterize the performance of a dipolar quantum battery system. These quantities are plotted as functions of time $t$ for different values of the dipolar coupling constants (both $\epsilon$ and $\Delta$) set to 2 (pink-solid), 3 (green-dot-dashed), 4 (blue-dotted), and 5 (red-dashed).  Figure (\ref{Energy factor}a) shows how the extractable work, or ergotropy $W(t)$, evolves over time. As the dipolar coupling increases, the amplitude of oscillations in $W(t)$ becomes larger, indicating stronger charging and discharging dynamics. A higher dipolar coupling leads to greater energy storage and more efficient work extraction. While the periodicity remains unchanged, systems with stronger dipolar interactions exhibit higher peak ergotropy, signifying enhanced battery performance. 
\begin{figure}[t]
	\centering
	\includegraphics[width=0.9\textwidth]{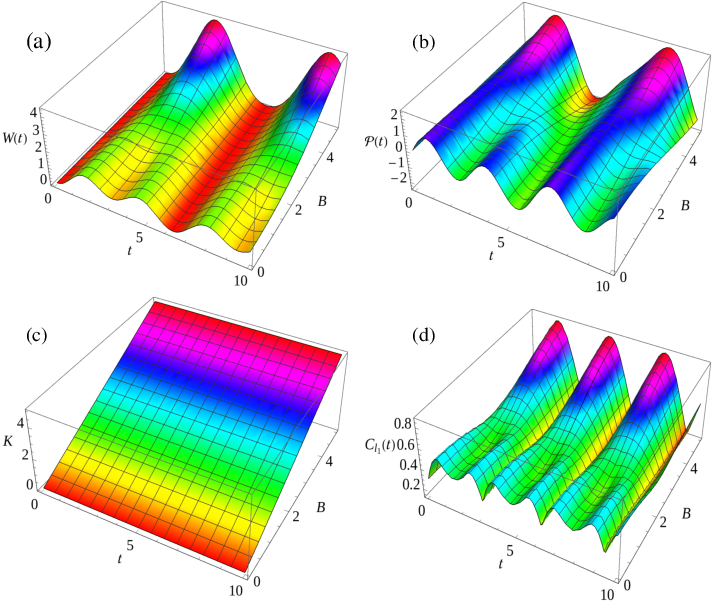}
	\caption{Evolution of (a) ergotropy $W$, (b) power $\mathcal{P}$, (c) capacity $\mathit{K}$, and (d) $l_{1}$-norm of coherence $C_{l_1}$ in the dipolar quantum battery for different magnetic field $B$ and time $t$. Other parametric values are fixed as $\epsilon=\Delta=2$, $D=1$, $T=0.5$, and $\Omega=1$.}
	\label{3D Plot}
\end{figure}
In Figure (\ref{Energy factor}b), we plot the instantaneous power $\mathcal{P}(t)$ for the same set of dipolar couplings. Similar to ergotropy, $\mathcal{P}(t)$ exhibits oscillatory behavior, alternating between positive (charging) and negative (discharging) values. The extrema of the power curves increase with higher dipolar interaction strength, reflecting faster energy exchange dynamics. As the coupling strength grows, the system becomes increasingly dynamic in energy flow. Figure (\ref{Energy factor}c) presents the battery capacity $\mathit{K}$  as a function of time. Unlike ergotropy and power, the capacity remains nearly constant over time for all values of dipolar coupling, though its magnitude increases with stronger interactions. 
Lastly, Figure (\ref{Energy factor}d) depicts the time-dependent  $l_1-$norm of coherence,  which quantifies quantum coherence in the battery. Coherence increases with dipolar strength and exhibits temporal fluctuations. Since both power and ergotropy improve with stronger dipolar interactions, the observed increase in coherence further supports the idea that quantum coherence is a critical resource for enhancing battery performance. These results underscore the role of coherence-driven mechanisms in optimizing the functionality of dipolar quantum batteries. These findings suggest that dipolar coupling plays a constructive role in designing more effective and efficient quantum energy storage systems.

Figure \ref{3D Plot} illustrates the time evolution of four critical performance indicators of a dipolar quantum battery such as ergotropy $W(t)$, instaneous power $\mathcal{P}(t)$, capacity $\mathit{K}$, and $l_1-$norm of coherence $C_{l_1}(t)$ for varying values of the external magnetic field strength $B$. In Figure (\ref{3D Plot}a), we observe that as $B$ increases ergotropy  $W(t)$ exhibits enhanced oscillations, indicating that more energy can be stored and retrieved. The growing amplitude of these oscillations with increasing $B$ suggests that stronger magnetic fields enable the battery to store and extract more energy during unitary evolution. This enhancement in ergotropy demonstrates that magnetic field tuning is an effective strategy for boosting quantum battery performance. Figure (\ref{3D Plot}b) presents the oscillatory behavior of the instantaneous power $\mathcal{P}(t)$, reflecting the periodic charging and discharging of the battery. Higher magnetic field strengths lead to more pronounced oscillations in both charging and discharging phases, indicating an increased energy transfer rate. Thus, the magnetic field not only enhances the energy capacity but also makes the battery more dynamically responsive. 
 
 Figure (\ref{3D Plot}c) shows the battery's capacity $\mathit{K}$,  which remains constant over time across all magnetic field values, in accordance with the theoretical model. However, the magnitude of the capacity increases with increasing magnetic field strength, indicating that stronger fields enlarge the energy gap between the ground and excited states, thereby improving the battery's theoretical maximum energy storage. Figure (\ref{3D Plot}d) explores the coherence behavior of the battery. The coherence metric $C_{l_1}(t)$ displays temporal oscillations and increases with the strength of the magnetic field. Since quantum coherence is a vital resource for quantum energy storage, its enhancement with increasing $B$ underscores a strong correlation between coherence, work capacity, and power. This reinforces the view that quantum coherence is a key factor in enhancing battery performance. 

In summary, the magnetic field $B$ plays a pivotal role in controlling and enhancing the functional attributes of the quantum battery. These results highlight magnetic field tuning as a promising approach for designing and optimizing high-performance quantum energy storage systems.

\section{Effect of dephasing environment on Dipolar QB}\label{dephasesec}
To assess the robustness of the dipolar spin-system quantum battery, we investigate its
dynamics under the influence of a pure-dephasing environment \cite{Refereearticle3}. For this purpose, we consider the Hamiltonian of the system as 
\begin{align}
\mathcal{H}=\mathcal{H}_{BC}+\mathcal{H}_E+\mathcal{H}_{\text{int}},
\end{align}
where $\mathcal{H}_{BC}$ is the Hamiltonian of the battery–charger system, given by 
\begin{align}
\mathcal{H}_{BC}=-\frac{1}{3}\,\vec{S_1} \cdot \overleftrightarrow{\mathcal{D}}\cdot \vec{S_2}
+ \vec{D}\cdot(\vec{S_1}\times \vec{S_2})
+ \hbar \omega_0 S_B^z+\hbar \omega_0 S_C^z.
\label{ham1}
\end{align}
Here, $\hbar \omega_{0}$ denotes the energy spacing of the qubit levels of both the charger and the battery, which coincides with the magnetic terms appearing in Equation (\ref{ham}). The both charger and battery qubit interacting via dipolar and DM interactions.  The Hamiltonian of the environment is $\cal{H}_{R}$. The interaction Hamiltonian between the battery-charger system and the environment is expressed as
\begin{align}
H_{int}=\hbar(\gamma_BS_B^z+\gamma_CS_C^z)R,
\end{align}
where $\gamma_B$ and $\gamma_C$ describe the decoherence parameters of battery and charger respectively and R denotes the Hermitian operator of the environment. The decoherence process suppresses the coherent exchange of excitation between the charger and the battery, leading to a damping of oscillations in both the ergotropy and instantaneous power. The dynamics of the open system can be described by the Lindblad-type master equation
\begin{align}
    \dot{\rho}(t) = -i[H,\rho(t)] 
    + \sum_{q=B,C} \gamma_q 
    \Big( \sigma_z^{(q)} \rho(t) \sigma_z^{(q)} - \rho(t) \Big),
\end{align}
where $H$ is the system Hamiltonian, $\sigma_z^{(q)}$ acts on qubit $q$, and $\Gamma_{\varphi}=\gamma_B+\gamma_C$ denotes the effective dephasing rate. Restricting to the single-excitation subspace $\{|01\rangle,|10\rangle\}$, the reduced
density matrix can be parametrized as
\begin{align}
    \rho_{\text{sub}}(t) =
    \begin{pmatrix}
    u(t) & v(t) \\
    v^*(t) & 1-u(t)
    \end{pmatrix},
\end{align}
where $u(t)$ is the excited-state population and $v(t)$ the coherence. The coupled equations of motion take the form
\begin{align}
    \dot{u}(t) &= 2\kappa\, \text{Im}[v(t)], \\
    \dot{v}(t) &= -i\kappa \big(1-2u(t)\big) - \Gamma_{\varphi} v(t),
\end{align}
with $\kappa=\tfrac{1}{6}\sqrt{\Delta^2+9D^2}$ denoting the effective coupling strength. Analytically, the population imbalance obeys a damped harmonic equation of the form
\begin{align}
    \ddot{z}(t) + \Gamma_{\varphi}\dot{z}(t) + (2\kappa)^2 z(t) = 0.
\end{align}
The resulting solutions describe damped Rabi-type oscillations of ergotropy,
\begin{align}
    W(t) = \kappa e^{-\Gamma_{\varphi} t/2}
    \sqrt{\Big(\cos(\omega t)+\tfrac{\Gamma_{\varphi}}{2\omega}\sin(\omega t)\Big)^2
    + \Big(\tfrac{2\kappa}{\omega}\Big)^2 \sin^2(\omega t)},
\end{align}
with $\omega=\sqrt{(2\kappa)^2 - (\Gamma_{\varphi}/2)^2}$. 
In the long-time limit, the system approaches an equipartition state with vanishing ergotropy, highlighting the detrimental impact of dephasing noise on the charging capability. This analysis clearly establishes the critical role of coherence in sustaining efficient energy storage in quantum batteries. 

\begin{figure}[t]
	\centering
	\includegraphics[width=0.9\textwidth]{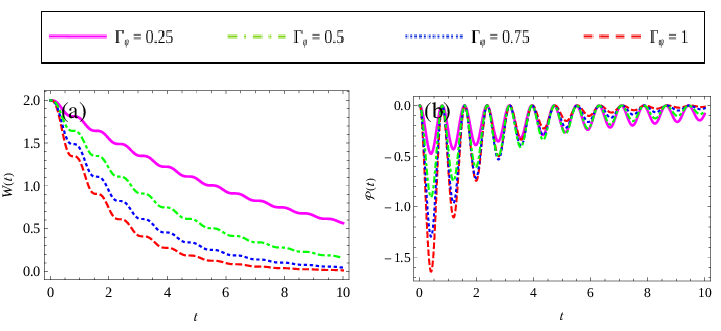}
	\caption{Time evolution of (a) ergotropy $W(t)$ and (b) power $\mathcal{P}(t)$ in the dipolar quantum battery for different decay rates $\Gamma_{\varphi}= 0.25, 0.5, 0.75,$ and $1$ in the absence of DM interaction ($D=0$). Other parameters are $\Delta = 2$ and $\Omega = 1$. The plots show that increasing $\Gamma_{\varphi}$ accelerates the decay of ergotropy and power, driving the system towards a passive state with no extractable work.}
	\label{Figdephase1}
\end{figure}

To demonstrate the performance of a dipolar quantum battery under the influence of battery–charger interaction in a dephasing environment, Figure \ref{Figdephase1} presents the ergotropy as a function of time for different decay rates $\Gamma_{\varphi}$, in the absence of DM interaction. As shown in Figure (\ref{Figdephase1}a), once the environmental effects via $\Gamma_{\varphi}$ are introduced, the ergotropy $W(t)$ of the battery–charger system begins to oscillate, indicating that the system can still store and release useful energy. However, as the decay rate $\Gamma_{\varphi}$ increases (for example, $\Gamma_{\varphi} = 0.5, 0.75,$ and $1$), the possibility of extracting maximum work from the battery decreases with time. In particular, at $\Gamma_{\varphi}=1$, the ergotropy $W(t)$ of the system rapidly decays to zero, implying that no useful work can be extracted. This behavior highlights that environmental decoherence destroys the system’s ability to sustain coherent superpositions, and since coherence is directly tied to work extraction, ergotropy eventually vanishes. In the absence of DM interaction, environmental noise dominates and drives the system into a fully mixed passive state, where no useful work can be obtained. The corresponding dynamics are confirmed by the time evolution of the power $\mathcal{P}(t)$, which exhibits damped oscillations during the transient regime and approaches zero as $t \rightarrow \infty$. As noted earlier, at $\Gamma_{\varphi}=1$, the power $\mathcal{P}(t)$ decays rapidly to zero. The negative values of power reflect the discharging dynamics of the battery–charger system in the presence of dephasing. Overall, Figure 6 highlights the vulnerability of quantum batteries: under realistic conditions, environmental decoherence strongly limits ergotropy and thus their practical usefulness.

\begin{figure}[t]
	\centering
	\includegraphics[width=0.9\textwidth]{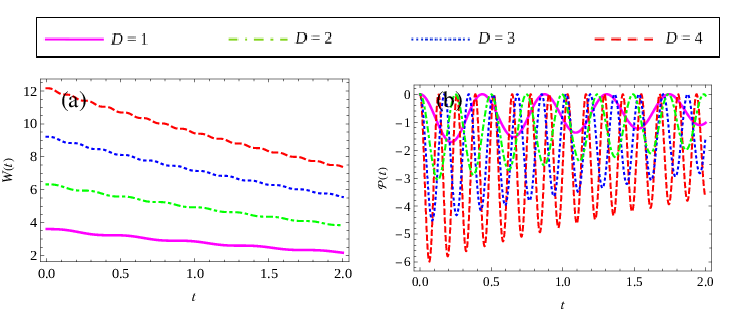}
	\caption{Time evolution of (a) ergotropy $W(t)$ and (b) power $\mathcal{P}(t)$ in the dipolar quantum battery for $\Gamma_{\varphi} = 0.5$ and different DM interaction strengths $D = 1, 2, 3,$ and $4$. Other parameters are $\Delta = 2$ and $\Omega = 1$. The results indicate that DM interaction enhances the initial ergotropy and power, enabling larger work extraction in the short-time regime, although both quantities decay to zero at long times due to dephasing.}
	\label{Figdephase2}
\end{figure}

Next, we explore the effect of the DM interaction (i.e., $D \neq 0$) on the performance of the battery–charger system in the presence of a dephasing environment. Figure \ref{Figdephase2} shows the time evolution of the battery's ergotropy $W(t)$ and power $\mathcal{P}(t)$ for $\Gamma_{\varphi}=0.5$ and different DM interaction strengths $D=1, 2, 3,$ and $4$. As discussed in Figure ~\ref{Fig2}, the DM interaction enhances the maximum value of $W(t)$ compared to the case without DM interaction ($D=0$). Indeed, the ergotropy reaches larger peak values as $D$ increases. However, the enhancement is only transient: as time evolves, $W(t)$ decays rapidly to zero, implying that the stored energy cannot be extracted at long times. This behavior arises because, although the DM interaction facilitates energy transfer within the system, the simultaneous coupling to the dephasing environment induces fluctuations that suppress the battery’s peak ergotropy. Figure (\ref{Figdephase2}b) further confirms this effect, showing oscillations of the power $\mathcal{P}(t)$ between zero and negative values. These negative regions reflect discharging dynamics, and ultimately, both $W(t)$ and $\mathcal{P}(t)$ decay to zero. Therefore, while the DM interaction initially improves energy extraction, the dephasing environment degrades the long-term performance of the dipolar quantum battery.

\begin{figure}[t]
	\centering
	\includegraphics[width=0.9\textwidth]{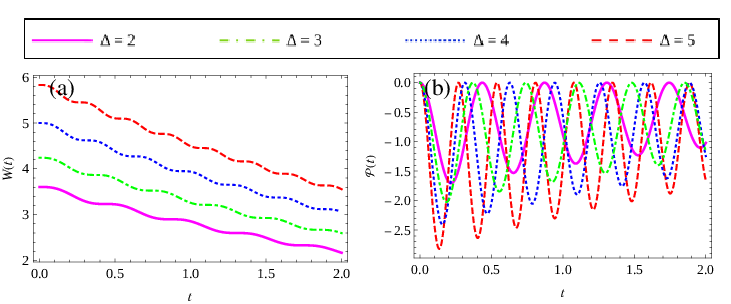}
	\caption{Time evolution of (a) ergotropy $W(t)$ and (b) power $\mathcal{P}(t)$ in the dipolar quantum battery for different dipolar coupling strengths $\Delta = 2, 3, 4,$ and $5$ in the presence of dephasing ($\Gamma_{\varphi} = 0.5$). Other parameters are $D = 1$ and $\Omega = 1$. The dipolar coupling modifies the oscillation amplitude, but dephasing dominates at long times, leading to vanishing ergotropy and power.}
	\label{Figdephase3}
\end{figure}

To illustrate the effects of dipolar coupling $\Delta$, the dynamics of the battery–charger system in the presence of a common dephasing environment are shown in Figure \ref{Figdephase3} for $\Delta = 2, 3, 4,$ and $5$. Similar to the previous case, both the ergotropy $W(t)$ and the power $\mathcal{P}(t)$ exhibit damped oscillations for all considered values of $\Delta$. As time evolves, these oscillations decay towards zero, reflecting the discharging dynamics induced by the dephasing environment. The results indicate that the influence of dipolar coupling is weaker than that of the dephasing process, which dominates and drives the system towards a fully mixed passive state. This observation is consistent with the findings in the previous subsection for different DM interaction strengths ($D = 1, 2, 3, 4$).

Collectively, these results emphasize that while internal interactions (DM and dipolar coupling) can transiently boost performance, environmental dephasing fundamentally limits the long-term work-extraction capability of dipolar quantum batteries.

\textit{Experimental Relavance} The model parameters admit a direct experimental interpretation. The dipolar tensor components $(\Delta,\epsilon)$ encode spectroscopically calibrated two-spin couplings; the DM vector magnitude $D$ captures spin–orbit-induced antisymmetric exchange measured in molecular dimers; the Zeeman term $B$ follows from the applied static field; the charging drive strength $\Omega$ is set by the RF/microwave Rabi frequency; and the pure-dephasing rate $\Gamma_\varphi$ reflects inhomogeneous broadening (e.g., spin–spin relaxation time in NMR/ESR). A practical comparison protocol is: (i) extract $\{\Delta,\epsilon,D,B,\Omega\}$ from standard spectroscopy and Rabi calibration, (ii) estimate $\Gamma_\varphi$ via echo or line-width analysis, and (iii) use these values in our closed-form/semi-analytic expressions for $W(t)$ and $\mathcal{P}(t)$ to predict charging/discharging transients. Agreement of peak positions, oscillation frequency, and damping envelope provides a quantitative benchmark of the model against real-device dynamics in NMR spin pairs \cite{NMRexp1, NMRexp2}, molecular magnets \cite{ Rydberg3}, or Rydberg dimers \cite{Rydberg1, Rydberg2}.
 
\section{Conclusion}
\label{concl}
In this work, we investigated the unitary dynamics of a dipolar quantum battery by analysing four key thermodynamic and quantum-informational metrics, in particular, ergotropy, instantaneous power, capacity, and the $l_{1}$-norm of coherence. Our two-qubit spin model, which incorporates both dipolar and Dzyaloshinskii–Moriya (DM) interactions, exhibits rich energy-storage behaviour that can be finely tuned through temperature, external magnetic field, and spin-interaction parameters. We find that lower temperatures and stronger DM coupling simultaneously enhance the battery’s extractable energy and quantum coherence, while magnetic-field tuning provides an additional lever to optimise both work extraction and coherence generation.

These results highlight the pivotal role of engineered spin–spin interactions and sustained quantum coherence in boosting the performance of quantum batteries, thereby charting a clear path toward their practical deployment in quantum-energy technologies.
 Additionally, our findings regarding the performance of the battery-charger system in a dephasing environment emphasize the necessity of properly considering and potentially engineering the inevitable existence of an external environment.

\noindent

\section*{Acknowledgment}
Authors are indebted to the referees for their critical comments to improve the contents of the manuscript.
V.K.C wish to thank DST, New Delhi for computational facilities under the DST-FIST program (SR/FST/PS-1/2020/135) to the
Department of Physics.

\end{document}